\documentclass[structabstract]{aa}  

\usepackage{graphicx}
\usepackage{txfonts}
\usepackage[normalem]{ulem}
\usepackage{comment}

\begin{document}

\title{He\,{\sc i} $\lambda$\,10830\,$\AA$ in the transmission spectrum of HD\,209458\,b}
%
%
\titlerunning{He\,{\sc i} in the transmission spectrum of HD\,209458\,b}
\author{F.~J. Alonso-Floriano\inst{1},  I.~A.~G. Snellen\inst{1}, S.~Czesla\inst{2},  F.~F.~Bauer\inst{3}, M.~Salz\inst{2},  M.~Lamp{\'o}n\inst{3}, L.~M.~Lara\inst{3}, E.~Nagel\inst{2}, M.~L\'opez-Puertas\inst{3}, L.~Nortmann\inst{4,5},  A.~S\'anchez-L\'opez\inst{3}, J.~Sanz-Forcada\inst{6}, J.~A.~Caballero\inst{6}, A.~Reiners\inst{7}, I.~Ribas\inst{8,9},  A.~Quirrenbach\inst{10}, P.~J. Amado\inst{2},  J.~Aceituno\inst{11}, G.~Anglada-Escud{\'e}\inst{12}, V.~J.~S.~B{\'e}jar\inst{5,6}, M.~Brinkm{\"o}ller\inst{10}, {A.~P.~Hatzes}\inst{13}, Th.~Henning\inst{14}, A.~Kaminski\inst{10}, M.~K{\"u}rster\inst{14}, F.~Labarga\inst{15}, D.~Montes\inst{15},  E.~Pall{\'e}\inst{5,6}, J.~H.~M.~M.~Schmitt\inst{3}, and M.~R.~Zapatero~Osorio\inst{16} 
}
\institute{Leiden Observatory, Leiden University, Postbus 9513, 2300 RA, Leiden, The Netherlands
\and
Hamburger Sternwarte, Universit{\"a}t Hamburg, Gojenbergsweg 112, 21029 Hamburg, Germany
\and
Instituto de Astrof{\'i}sica de Andaluc{\'i}a (IAA-CSIC), Glorieta de la Astronom{\'i}a s/n, 18008 Granada, Spain
\and
Instituto de Astrof{\'i}sica de Canarias (IAC), Calle V{\'i}a Lactea s/n, E-38200 La Laguna, Tenerife, Spain
\and
Departamento de Astrof{\'i}sica, Universidad de La Laguna, 38026  La Laguna, Tenerife, Spain
\and
Centro de Astrobiolog{\'i}a (CSIC-INTA), ESAC, Camino bajo del castillo s/n, 28692 Villanueva de la Ca{\~n}ada, Madrid, Spain
\and
Institut f{\"u}r Astrophysik, Georg-August-Universit{\"a}t, 37077 G{\"o}ttingen, Germany
\and
Institut de Ci\`encies de l'Espai (CSIC-IEEC), Campus UAB, c/ de Can Magrans s/n, 08193 Bellaterra, Barcelona, Spain
\and
Institut d'Estudis Espacials de Catalunya (IEEC), 08034 Barcelona, Spain
\and
Landessternwarte, Zentrum f\"ur Astronomie der Universit\"at Heidelberg, K\"onigstuhl 12, 69117 Heidelberg, Germany
\and
Centro Astron{\'o}nomico Hispano Alem{\'a}n, Observatorio de Calar Alto, Sierra de los Filabres, E-04550 G{\'e}rgal, Spain
\and
School of Physics and Astronomy, Queen Mary, University of London, 327 Mile End Road, London, E1 4NS, UK
\and
Th{\"u}ringer Landessternwarte Tautenburg, Sternwarte 5, 07778 Tautenburg, Germany
\and
Max-Planck-Institut f{\"u}r Astronomie, K{\"o}nigstuhl 17, 69117 Heidelberg, Germany
\and
{Departamento de F\'{i}sica de la Tierra y Astrof\'{i}sica 
and IPARCOS-UCM (Intituto de F\'{i}sica de Part\'{i}culas y del Cosmos de la UCM), Facultad de Ciencias F\'{i}sicas, Universidad Complutense de Madrid, E-28040, Madrid, Spain}
\and
Centro de Astrobiolog{\'i}a {(CSIC-INTA)}, Carretera de Ajalvir km 4, E-28850 Torrej{\'o}n de Ardoz, Madrid, Spain
}
\authorrunning{F.~J. Alonso-Floriano et al.}
\date{Received 29 May 2019; Accepted 30 July 2019}

\abstract
{Recently, the He\,{\sc i} triplet at 10830\,$\AA$ has been rediscovered as an excellent probe of the extended and possibly evaporating atmospheres of close-in transiting planets. This has already resulted in detections of {this triplet in the atmospheres of} a handful of planets, both from space and from the ground. However, while a strong signal is expected for the hot Jupiter HD\,209458\,b, only upper {limits have} been obtained so far.}
{Our goal is to measure the {helium excess absorption} from HD\,209458\,b and assess {the} extended atmosphere {of the planet} and possible evaporation.}
{We obtained {new} high-resolution spectral transit time-series of {HD\,209458\,b} using CARMENES at the 3.5\,m Calar Alto telescope, targeting the He\,{\sc i} triplet at 10830\,$\AA$ at a spectral resolving power of 80\,400. {The observed} spectra were corrected for stellar absorption lines using out of transit data, for telluric absorption using the {\sc molecfit} software, and for the sky emission lines using simultaneous sky measurements through a second fibre.}
{We detect He\,{\sc i} absorption at a level of 0.91\,$\pm$\,0.10\% {(9\,$\sigma$)} at mid-transit. {The absorption follows the radial velocity change of the planet during transit, unambiguously identifying the planet as the source of the absorption.} The core of the absorption exhibits a net blueshift of 1.8\,$\pm$\,1.3\,km\,s$^{-1}$.  Possible low-level excess absorption is seen further blueward from the main absorption near the centre of the transit, which could {be caused by} an extended tail. However, this needs to be confirmed.
}
{Our results further support a close relation between the strength of planetary absorption in the helium triplet lines and the level of ionising, stellar {X-ray and extreme-UV} irradiation.}
\maketitle
%

\section{Introduction}
\label{section.introduction}

Atmospheric mass loss can have a profound influence on the evolution of a planet, including the early evolution of the Earth \citep{Lammer2008} and other Solar System planets. Hot Jupiters, gas giants in {tight orbits}, are ideal objects to study atmospheric mass loss in detail. A large, extended atmosphere was first detected
around HD\,209458\,b targeting {the} Ly$\alpha$ line \citep{Vidal-Madjar2003}, showing a deep  transit signal pointing at atmospheric escape. While the {extended atmosphere} around this planet was subsequently also detected in carbon and oxygen \citep{Vidal-Madjar2004}, {Ly$\alpha$ has remained a primary tool to study planetary atmospheres and their escape} (e.g., { \citealt{Lecavelier2010,Kulow2014,Ehrenreich2015,Lavie2017,Bourrier2017,Bourrier2018}}). Most notably, \cite{Ehrenreich2015} {and later \cite{Lavie2017}} showed that the hot Neptune GJ\,436\,b is surrounded by a giant comet-like cloud of hydrogen, extending far beyond the Roche radius. 

Unfortunately, Ly$\alpha$ is located in the ultraviolet part of the spectrum and currently only accessible using the {\it Hubble Space Telescope} ({\it HST}). Furthermore, observations are strongly hampered by {interstellar absorption} and air glow emission originating from a halo of hydrogen atoms around the Earth, contaminating the profile of the Ly$\alpha$ line. 
Almost two decades ago, \cite{Seager&Sasselov2000} proposed the He\,{\sc i} transition at 10830\,$\AA$ {(in air)}, {a triplet of absorption lines}  of  a  metastable  state  of  helium, as a good candidate to study the extended atmospheres of close-in giant planets. \cite{Moutou2003} studied this helium transition during a transit of HD\,209458\,b {– observed with the spectroscopic
mode of ISAAC on the VLT –} and obtained a 3\,$\sigma$ upper limit of 0.5\% for a 3\,$\AA$ bandwidth. 

Recently, three independent groups detected He\,{\sc i}\,$\lambda$\,10830\,$\AA$ in exoplanet atmospheres. \cite{Spake2018} detected the triplet in WASP-107\,b using
spectrophotometry with \textit{HST}/WFC3. Spectroscopic detections at high-resolution were made in WASP-69 b, HAT-P-11\,b, and HD\,189733~b \citep[{of 3.59\,$\pm$\,0.19\%, $\sim$1.2\,$\pm$\,0.2\%}\footnote{{The error is an approximation obtained from the average transmission spectrum provided by \cite{Allart2018}.}}{, and 1.04\,$\pm$\,0.09\%, maximum reached absorptions respectively}]{Nortmann2018,Allart2018,Salz2018}. Another spectrophotometrical detection with \textit{HST}/WFC3 was later made in HAT-P-11\,b \citep{Mansfield2018} and ground spectroscopic observations of WASP-107\,b \citep[{7.92\,$\pm$\,1.00\%}]{Allart2019} confirmed the earlier $HST$ detection {and revealed detailed information on the helium escape as in the form of a cometary-like tail}.

The ground-based measurements are particularly informative, since they are conducted using high-dispersion spectroscopy -- in all four cases with the CARMENES spectrograph on the 3.5m Calar-Alto Telescope \citep{Quirrenbach2016,Quirrenbach2018}. At a resolving power of 80\,400, details on the radial velocity distribution of the helium gas are available. For example, \cite{Nortmann2018} measured blueshifts of several kilometers per second, which in combination with post-transit absorption, was interpreted as the escape of part of the atmosphere trailing behind the planet in comet-like form. 

\cite{Nortmann2018} also provided He\,{\sc i} upper limits for three additional planets, HD\,209458\,b, Kelt-9\,b, and GJ\,436\,b, of 0.84\%, 0.33\%, and 0.41\%, respectively (90\% confidence limits). 
The wide range of He\,{\sc i} absorption levels found for close-in planets is likely linked to the level of stellar X-ray and extreme-UV {(5--504\,\AA)} irradiation  of the host star, {populating the metastable} 2$^3$S helium triple state \citep{Nortmann2018}. {A similar relation between the He triplet and ionizing XUV ({$\lambda$}\,<\,504\,\AA) radiation was previously observed in stellar coronae \citep[e.g.,][]{Sanz-Forcada&Dupree2008}.} 

The observations {presented in \cite{Nortmann2018}} for HD\,209458\,b were performed under sub-optimal conditions, {resulting in poor data quality} that hampered the search for helium {(see Sect.\,\ref{sect.discussion})}. In this work, we present new CARMENES transit observations targeting the He\,{\sc i}$\,\lambda$\,10830\,$\AA$ triplet, now resulting in a firm detection at 0.91\,$\pm$\,0.10\%. In Section~2 we present the observations, in Section~3 the data analysis, in Section~4 the discussion on the results, and we conclude in Section~5.



\begin{table}
\centering
\caption{\label{table.parameters}Parameters of the exoplanet system HD\,209458.}
\begin{tabular}{l c c} 
   \hline
   \hline
   \noalign{\smallskip}
Parameter					 & Value & Reference$^{c}$  \\
  \noalign{\smallskip}
    \hline
      \noalign{\smallskip}
$\alpha$ [J2000]	& 22:03:10.77 	& 1	\\
\noalign{\smallskip}
$\delta$ [J2000]	& +18:53:03.5	& 1 \\
\noalign{\smallskip}
{$d$}         &   {48.37\,(12)\,pc}  & 1 \\ 
\noalign{\smallskip}
$V$			& {7.63\,(1)\,mag}	&	2 \\
\noalign{\smallskip}
$J$			& {6.59\,(2)\,mag}	&	3 \\
\noalign{\smallskip}
$\varv_{\rm sys}$ 	& --14.7652\,(16)\,km\,s$^{-1}$ & 4 	\\
\noalign{\smallskip}
$K_{\star}$ 		& {84.67\,(70)\,m\,s$^{-1}$}	& 5 	\\
\noalign{\smallskip}
$R_{\star}$		& 1.155$^{+0.014}_{-0.016}$\,$R_\sun$& 5 	\\
\noalign{\smallskip}
$M_{\star}$		& {1.119\,(33)\,$M_\sun$} & 5	\\
\noalign{\smallskip}
$T_{\rm eff}$ 		& {6065\,(50)\,K}	& 5	\\
\noalign{\smallskip}
$F_{XUV}^{a}$ & {1.004\,(284)\,W\,m$^{-2}$} & 6  \\
\noalign{\smallskip}
\noalign{\smallskip}
    \hline
\noalign{\smallskip}
${P_{\rm orb}}$ & 3.52474859\,(38)\,d & 7 \\
\noalign{\smallskip}
$T_{\rm 0}$\,[HJD] & 2452826.628521\,(87)\,d & 7  \\
\noalign{\smallskip}
{\it t}${_d}$ & {183.89\,(3.17)\,min} & 8 \\
\noalign{\smallskip}
{\em i} & {86.71\,(5)\,deg} & 5  \\
\noalign{\smallskip}
$a$		& 0.04707$^{+0.00046}_{-0.00047}$\,au	& 5 \\
\noalign{\smallskip}
$e\cos{\omega}$		& 0.00004\,(33) & 9  \\
\noalign{\smallskip}
$R_{\rm P}$ 	& 1.359\,$^{+0.016}_{-0.019}$\,$R_{J}$	& 5	\\
\noalign{\smallskip}
$M_{\rm P}$ 	& 0.685\,$^{+0.015}_{-0.014}$\,$M_{J}$	& 5	\\
\noalign{\smallskip}
$g_{\rm P}$ 	& {9.18\,(1)\,m\,s$^{-2}$}	& 5	\\
\noalign{\smallskip}
$T_{\rm eq}$ & {1449\,(12)\,K} & 5   \\
\noalign{\smallskip}
$K_{\rm P}^{b}$ 		& {144.9$^{+5.4}_{-5.3}$\,km\,s$^{-1}$}	&  6	\\
\noalign{\smallskip}
\hline
\end{tabular}
\tablefoot{
\tablefoottext{a}{X-ray and EUV (5--504\,\AA) flux at the distance of the planet's semi-major axis, derived from coronal models}.
\tablefoottext{b}{Derived from orbital parameters.}
\tablefoottext{c}{{References. (1)\,\citet{Gaia2018}, (2)\,\citet{Hog2000}, (3)\,\citet{Skrutskie2006}, (4)\,\citet{Mazeh2000}, (5)\,\citet{Torres2008}, (6)\,This work, (7)\,\citet{Knutson2007}, (8)\,\citet{Wright2011}, (9)\,\citet{Crossfield2012}.}}}
\end{table}

\section{Observations} \label{section.observations}

We observed the system HD\,209458 with CARMENES on 5 September 2018\footnote{{Program H18-3.5-022, P.I. S.~Czesla.} After one year from the observation date, the reduced spectra can be downloaded from the Calar Alto archive, {\tt http://caha.sdc.cab.inta-csic.es/calto/}}.
The planet is an archetypal hot Jupiter, and the first known transiting system of its kind (\citealt{Charbonneau2000,Henry2000}; Table\,\ref{table.parameters}). As a bright nearby system, it has been the subject of many observational studies targeting its transmission spectrum \citep[e.g.][]{Charbonneau2002,Deming2005b,Snellen2008,Snellen2010,Hoeijmakers2015, Hawker2018}, its hydrogen {upper atmosphere} \citep{Vidal-Madjar2003,Ben-Jaffel2007, Ehrenreich2008}, and its emission spectrum \citep{Deming2005a,Knutson2008,Swain2009,Schwarz2015,Brogi2017}.
This makes it one of the best studied exoplanets and a benchmark for atmospheric evaporation modelling. 

\begin{figure}
\centering
\includegraphics[angle= 0, width=0.49\textwidth,trim={0 200 0 200}, clip, keepaspectratio]{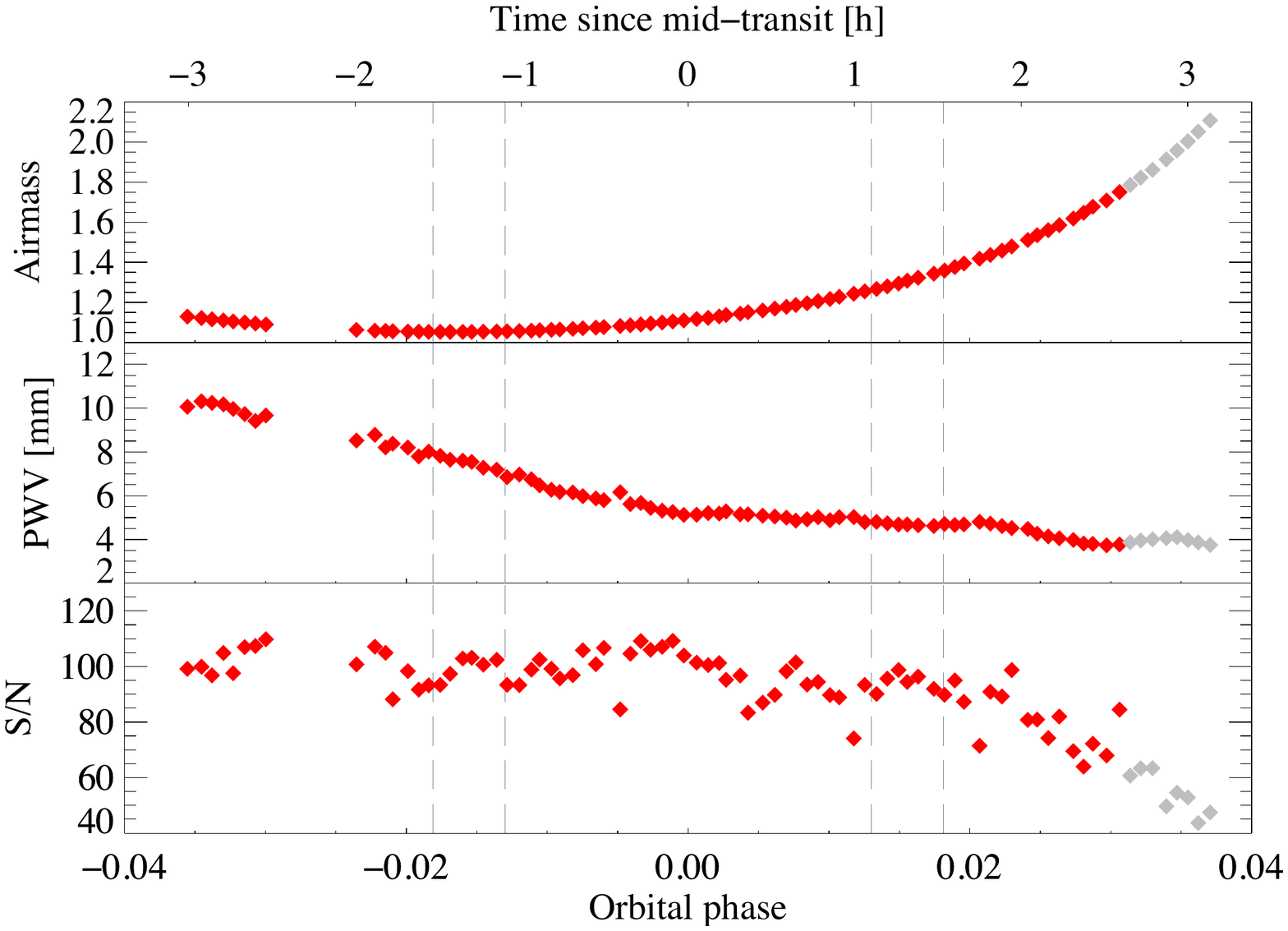}
\caption{\label{fig.observation_conditions} Change in airmass {({\em top})}, precipitable water vapour {({\em middle})}, and signal-to-noise ratio {({\em bottom})} during observations (mid-transit $\sim$\,00:39\,UT). {Vertical dashed lines indicate the first, second, third, and fourth contact, respectively.} {The lack of data around $-$2.3\,h was due to a lost in the tracking system}. The post-transit observations not included in the analysis are shown in grey.  {The out-of-transit red diamonds were used to compute the master spectrum of Fig.\,\ref{fig.master_spectrum}.}}
\end{figure}

CARMENES consists of two spectrograph channels fed by fibres connected to the front-end mounted on the telescope. One of the CARMENES spectrographs,  dubbed VIS channel, covers the optical wavelength range  $\Delta\lambda$\,=\,520--960\,nm over 55 orders. The other, the NIR channel, covers the near-infrared wavelength range $\Delta\lambda$\,=\,960--1710\,nm over 28 orders. The resolving power is $\mathcal{R}$\,=\,94\,600 in the VIS channel and $\mathcal{R}$\,=\,80\,400 in the NIR channel.  {Two fibres are connected to each channel: fibre A is used for the target, and fibre B for the sky. }
The observations were obtained in service mode and {consisted} of 91 exposures of 198\,s, starting at 21:39 UT and ending at 03:47 UT on the night starting {on} 5 September 2018, corresponding to a planet orbital phase range of {$\phi$\,=\,--0.036} to +0.037. {The tracking of the target was lost} during the pre-transit phase from 22:12 UT to 22:39 UT ({$\phi$\,=\,--0.029 to --0.024}, Fig.\,\ref{fig.observation_conditions}). We {did} not include these spectra in our analysis.
A typical continuum signal-to-noise ratio (S/N) of $\sim$95 per spectrum was reached around the He\,{\sc i} triplet. Since the quality of the data decreased significantly during the post-transit, down to a S/N of $\sim$40--60, we also discarded the last eight spectra (from  03:14 UT onward).

{During the observing run, the airmass ranged between 1.05 and 2.11}. We regularly updated the atmospheric dispersion corrector, requiring a re-acquisition of the target, each time taking about two minutes \citep[c.f.][]{Seifert2012}. {The update was done every $\sim$\,40\,min when the target was observed at altitudes >70$^{\circ}$ (airmass <1.06), every $\sim$\,30\,min at altitudes between 70$^{\circ}$ and 50$^{\circ}$ (airmass 1.06--1.30), and every $\sim$15\,min at lower altitudes (airmass >1.30).}
During the course of the observation, {the column} of precipitable water vapour (PWV) towards the target {(measured from the spectra)} decreased from 10.3 to 3.8\,mm.

\section{Data analysis and results}
\label{section.analysis}

The observed spectra were reduced using the CARMENES pipeline {\sc Caracal} v2.10 \citep{Zechmeister2014,Caballero2016}. The pipeline provides a vacuum wavelength solution, which we converted into air wavelengths, as used in the remainder of the paper. 

The standard bad pixels mask {of the NIR detector} included in {\sc Caracal} v2.10 {did} not sufficiently {correct} for hot pixel effects. In the data of fibre\,B, we included {in the mask} the left and right neighbours around the hot pixels {in the dispersion direction of the detector}, while in fibre\,A {we also included} the top and bottom neighbours. {A pair of remaining bad pixels present in the extracted spectra were removed manually (pixels 482 and 489 corresponding to 10829.357 and 10829.757\,$\AA$)}. In one of the observed spectra, taken at 00:13:58\,UT, irregular variations in the continuum were recorded {for some orders}. However, {they did} not affect the spectral order containing the He\,{\sc i} triplet.  

\begin{figure}
\centering
\includegraphics[angle= 0, width=0.49\textwidth,,trim={10 200 10 200}, clip, keepaspectratio]{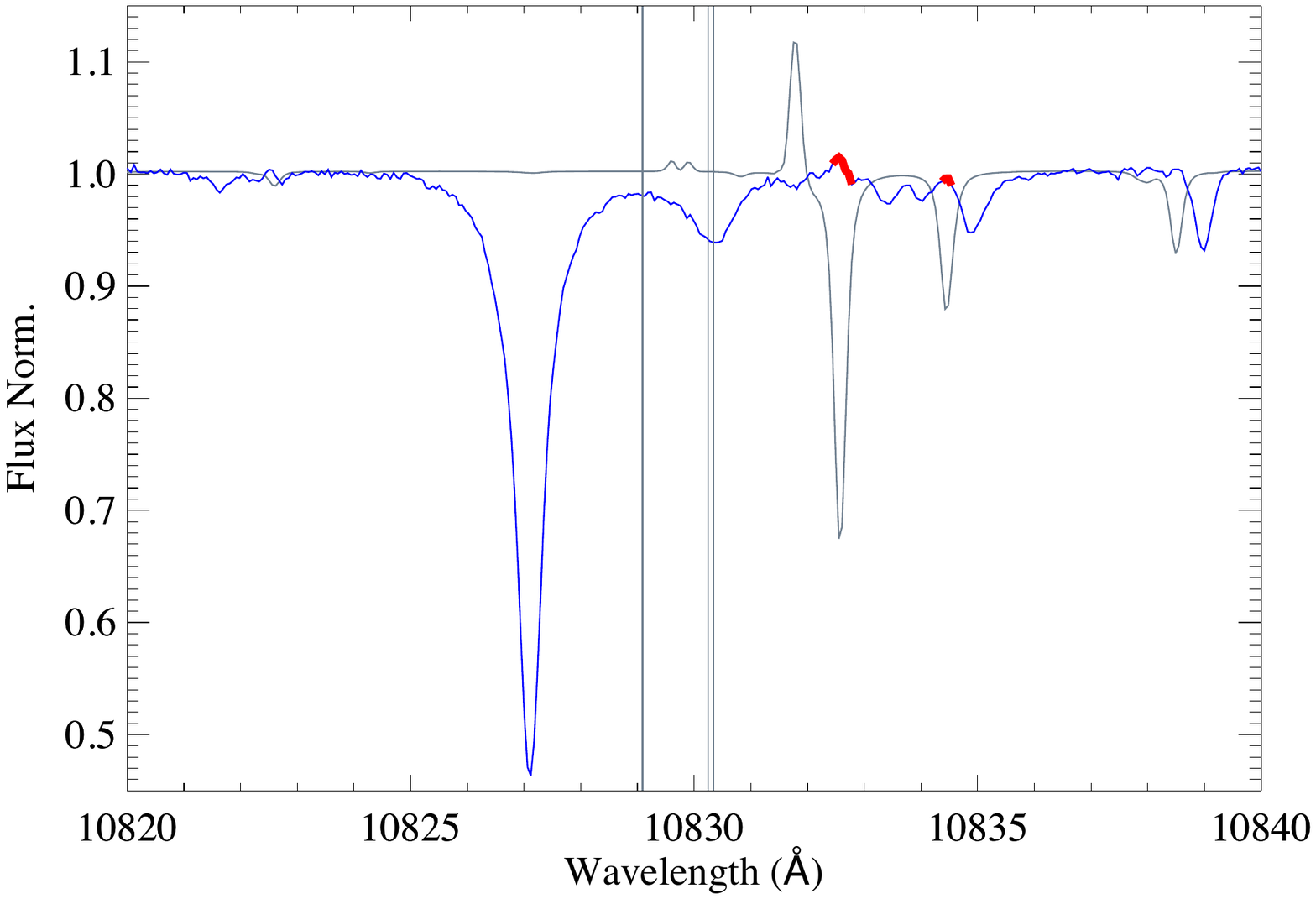}

\caption{\label{fig.master_spectrum}
The master spectrum in the vicinity of the He {\sc i} triplet (blue line). The vertical lines indicate the positions of the three He {\sc i} lines. The grey line indicates the average telluric absorption spectrum and the sky emission lines removed from the data. {The masked cores of telluric absorption lines in Fig.\,\ref{fig.contour_he} are indicated by thicker red lines.}{ The latter were highly variable during the night}. The stellar line blueward from the He triplet is Si\,{\sc i}.}
\end{figure}

\begin{figure}

\includegraphics[angle= 0, width=0.49\textwidth,,trim={0 200 0 200}, clip, keepaspectratio]{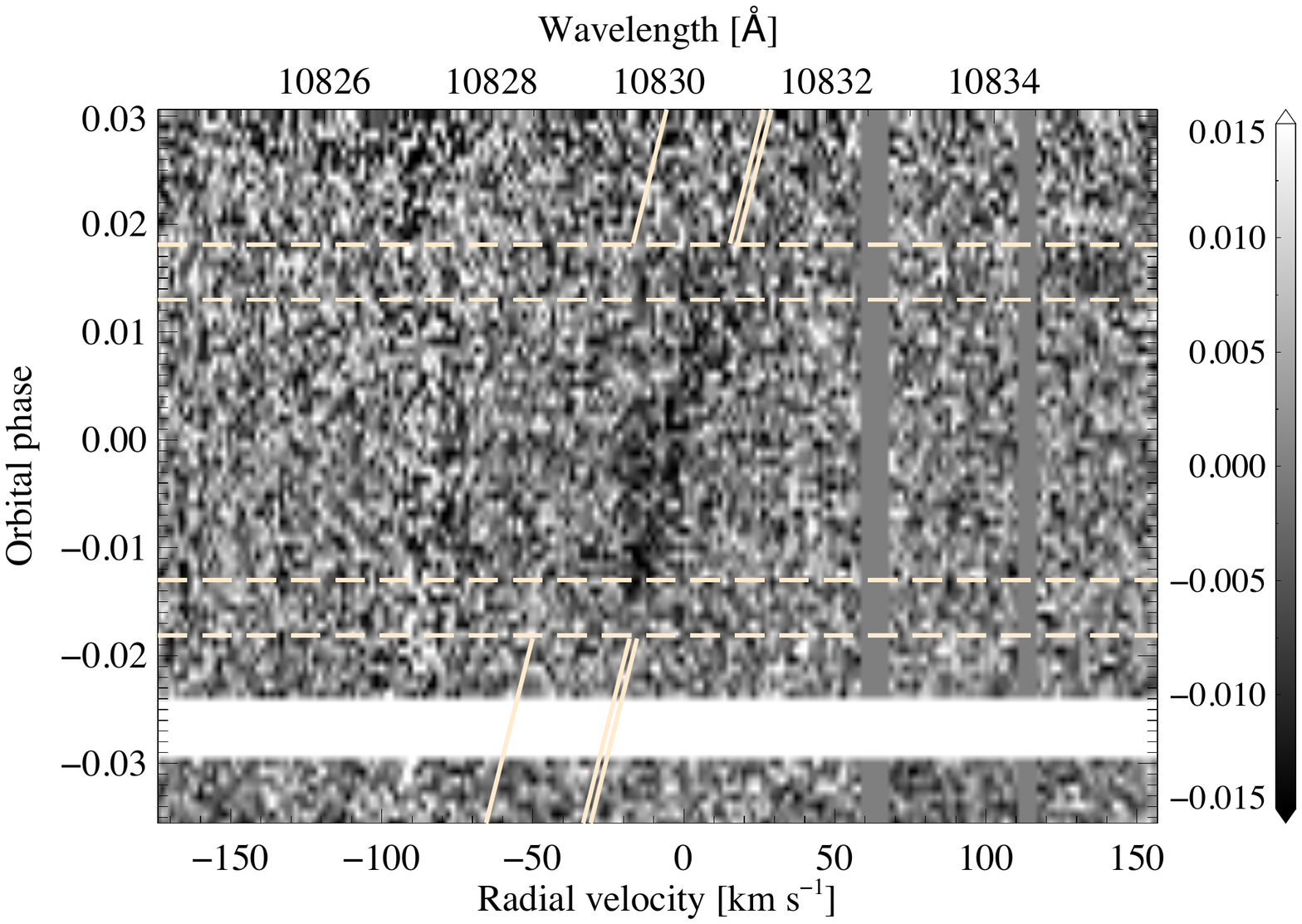}
\caption{\label{fig.contour_he}
Time sequence of the {residual} spectra around the He {\sc i} triplet, after removal of the stellar and telluric absorption and the sky emission lines, with the relative flux indicated in grey scale. The radial velocity in the stellar rest-frame is on the horizontal axis, and orbital phase on the vertical axis. The corresponding wavelengths in the rest-frame of the star are indicated in the top axis. 
The area around orbital phase $-$0.03 is blanked out due to {the lack of observations}. The horizontal dashed lines indicate {the first, second, third and fourth contact of the transit, respectively}. A helium signal is visible (darker grey scales), following the planet velocity ({slanted lines}) from $-$16 to $+$16\,km\,s$^{-1}$.}
\end{figure}

Subsequently, a similar process as in \cite{Nortmann2018} and \cite{Salz2018} was followed to remove the stellar and telluric absorption and {the sky emission lines} from the spectra around the He\,{\sc i} triplet. 
We used version 1.5.9 of the {\sc molecfit} software \citep{Smette2015, Kausch2015} to remove the telluric absorption lines. However, {as explained by \cite{Shulyak2019}, the telluric removal depends strongly on the S/N of the data
and atmospheric humidity, which might cause artefacts in the core of {the telluric} lines after correction. Therefore, the cores of the {telluric} lines could not be adequately corrected, and these regions were masked out (see Figs.\,\ref{fig.master_spectrum} and \ref{fig.contour_he})}. We subtracted the sky emission lines using the sky spectra from fibre B {as explained in \cite{Salz2018}}.
The fibre B spectra were {extracted in the same way than fibre A and} corrected for cosmic rays by fitting the temporal variation of each wavelength {element} with a high-order polynomial, and substituting those values that deviate {by more than} 5\,$\sigma$. 

We {then} normalised the {stellar} spectra by fitting their continuum, removing the cosmic-rays as in fibre B, and shifting the spectra to the stellar rest-frame using the barycentric and systemic velocities. A master spectrum was created (Fig.\,\ref{fig.master_spectrum}) combining the out-of-transit spectra with weights at each wavelength step based on the S/N following \( w_{\lambda}^{i}= x_{\lambda}^{i\,2}/\sum_{i} x_{\lambda}^{i\,2} \), 
where \(x_{\lambda}\) is the S/N at the wavelength, and {\it i} is the consecutive number of the spectrum. 
Subsequently, each {normalised} spectrum was divided by the master spectrum, resulting in an array of residuals (wavelength vs. time) shown in Fig.\,\ref{fig.contour_he}. 

\subsection{Transmission spectrum and light curve}

Figure\,\ref{fig.contour_he} shows the time sequence of {the residual} spectra around the He {\sc i} triplet after removal of the stellar and telluric absorption and the sky emission lines.  {{A helium signal} is visible during transit following the planet radial velocity}. This provides strong evidence that the helium absorption has a planetary origin. In the spectra obtained during mid-transit, a tentative extra absorption feature is visible directly blueward from the main helium absorption, which we discuss below.
The vertical features redward of He\,{\sc i} (60 and 120\,km\,s$^{-1}$) are masked areas corresponding to the centres of strong, highly variable telluric lines.
Residuals blueward ($-$80\,km\,s$^{-1}$) from He\,{\sc i}  are from stellar Si\,{\sc i} (Fig.\,\ref{fig.master_spectrum}) and {possibly} caused by the Rossiter-McLaughlin effect {(see Sect.\,\ref{subsect.line_profile})}.

\begin{figure}
\centering
\includegraphics[angle= 0, width=0.49\textwidth,trim={10 200 10 200}, clip, keepaspectratio]{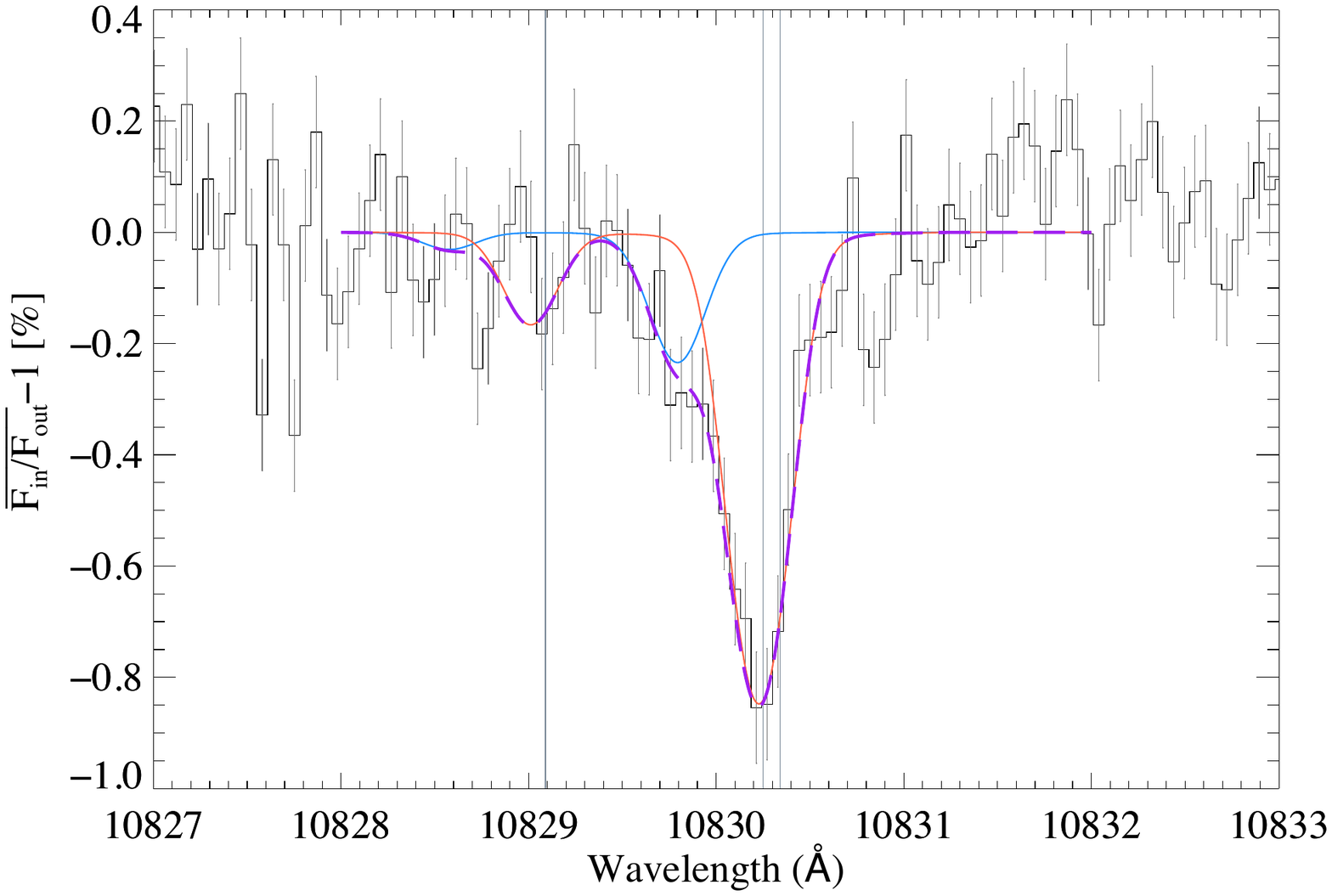}
\caption{\label{fig.transmission_spectrum}
Average transmission spectrum around the He\,{\sc i} triplet in the planet rest-frame. The positions of the three helium lines are marked by vertical lines. The red curve is the best-fit Parker wind model obtained for a temperature of 6000\,K and a mass-loss rate of 4.2$\times 10^{9}$\,g\,s$^{-1}$. The blue line indicates a model for a tentative blueward component centred at around $-$13\,km\,s$^{-1}$.  
{{The magenta dashed} curve is the combination of the red and blue lines.}}
\end{figure}

We {shifted the residual spectra into the planet rest-frame and }computed the average transmission spectrum of HD\,209458\,b using the 33 spectra collected between the second and third contact of the transit ($\phi$\,=\,--0.013 to +0.013, Fig.\,\ref{fig.transmission_spectrum}).
{The peak value of the average absorption signal in the core of the two
strongest and blended lines of the He\,{\sc i} triplet was at the level of $0.91\,\pm\,0.10\,\%$. The average absorption level over a bandwidth of 0.30\,$\AA$ centred at the absorption peak was $0.71\,\pm\,0.06\,\%$. In both cases, the absorption values and uncertainties were determined using the bootstrap method (Fig.\,\ref{fig.histogram}) as in \cite{Salz2018}.}
The third and weakest line of the triplet is not detected {(see below)}. 
The shape of the signal shows an asymmetry on the blue side, which can also be seen at mid-transit in the two-dimensional array of Fig.\,\ref{fig.contour_he}. {We further discuss the transmission line profile in Sect.\,\ref{subsect.line_profile}}.

We measure the peak of the helium absorption to be blueshifted by 1.8\,$\pm$\,1.3\,km\,s$^{-1}$ with respect to the restframe of the planet. This is compatible with the blueshift of 2\,$\pm$\,1\,km\,s$^{-1}$ observed for carbon monoxide by \cite{Snellen2010}. However, this absorption is likely originated from a different layer in the atmosphere of the planet.

\begin{figure}
\centering
\includegraphics[angle= 0, width=0.49\textwidth,,trim={10 200 20 200}, clip, keepaspectratio]{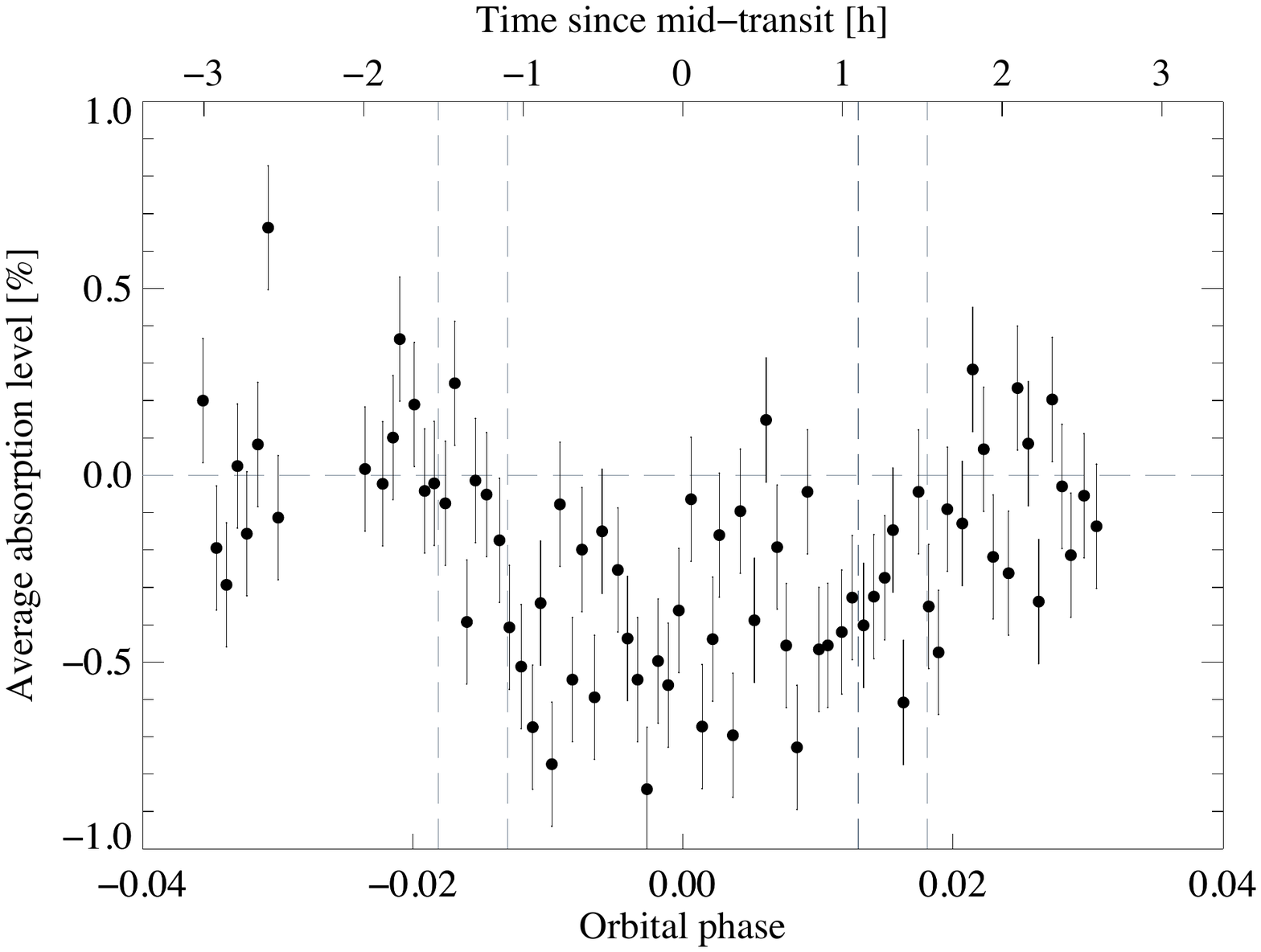}
\caption{\label{fig.light_curve}
Light curve of the He\,{\sc i} absorption centred on the observed core of the line and using a width of $\Delta\lambda\sim$0.9\,$\AA$ (total $\Delta v$\,=\,25.3\,km\,s$^{-1}$). Vertical dashed lines indicate the first, second, third, and fourth contact, respectively. There is no evidence for any out of transit absorption.
}
\end{figure}

We constructed the light curve of the He\,{\sc i} signal by measuring the average absorption per spectrum between --13.3 and +12.0\,km\,s$^{-1}$ (i.e., between 10829.814 and 10830.729\,$\AA$) in the planet rest-frame, which is shown in  Fig.\,\ref{fig.light_curve}. The average in-transit absorption signal was $\sim$0.44\%, about a factor two smaller than the peak transmission signal as shown in Fig.\,\ref{fig.transmission_spectrum} due to the relatively wide integration band. There is no evidence for a pre- or post-transit absorption signal. 

\begin{figure}
\centering
\includegraphics[angle= 0, width=0.49\textwidth,keepaspectratio,trim={40 0 20 20, clip}]{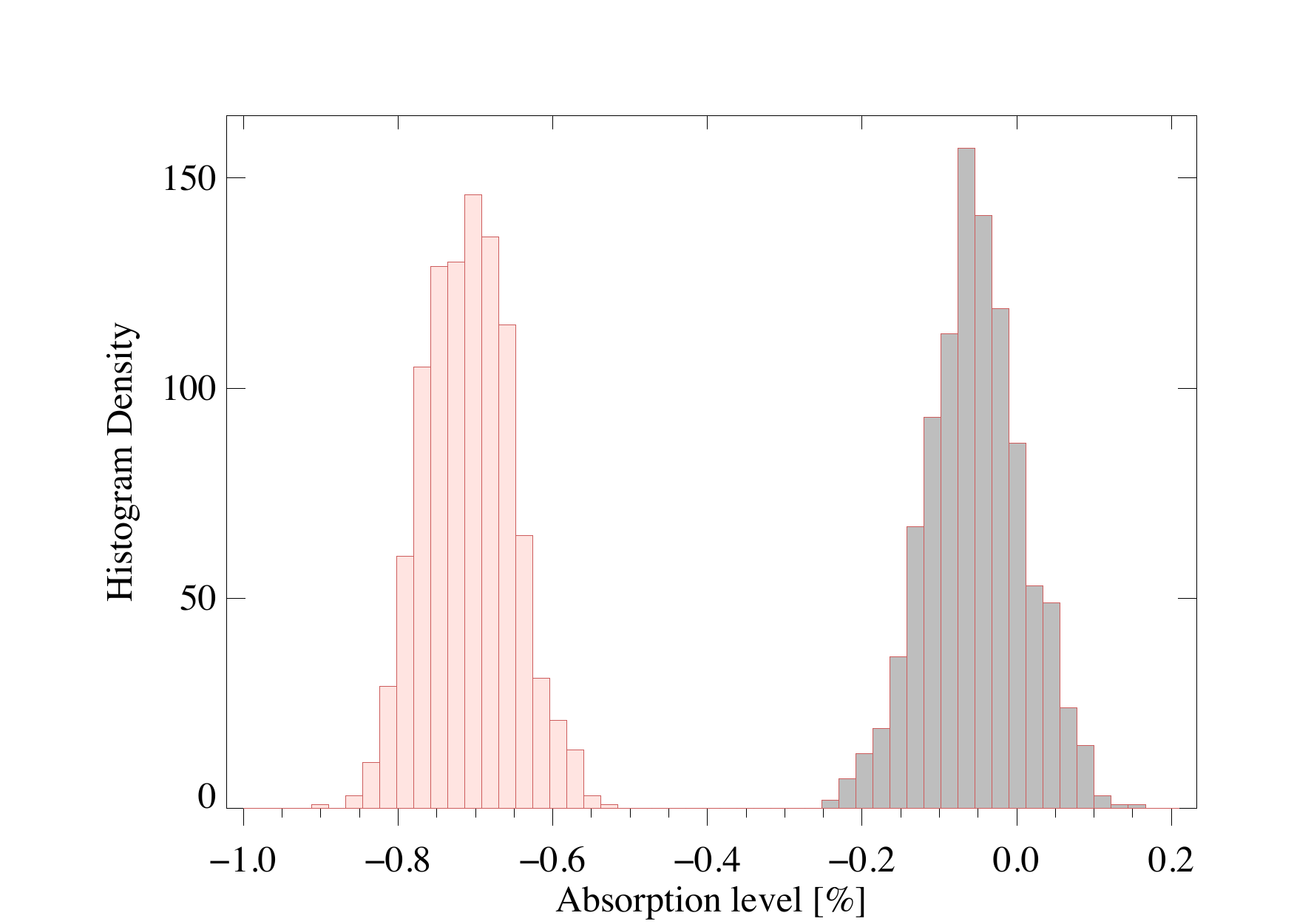}
\caption{\label{fig.histogram}
{Histograms of the mean absorption levels in the average transmission spectra generated in our bootstrap analysis when considering only spectra between the second and third contact ({\em left}) and a control sample including only the out-of-transit observations ({\em right}).}}

\end{figure}

\subsection{Modelling of the helium triplet absorption}
\label{subsect.modelling}

{A one-dimensional isothermal hydrodynamic and spherically symmetric model was used to calculate the He\,2$^{3}S$ density in the planet upper atmosphere, similar {to} that developed by \cite{Oklopcic&Hirata2018}. For a given range of temperatures and mass-loss rates {($\dot{M}$, where it refers to the total hydrogen and helium mass loss)}, the radial density and velocity profiles were computed by means of an isothermal Parker wind model \citep{Parker1958}. Afterwards, the continuity equations were solved to derive the He\,2$^{3}S$ density profile. The He triplet absorption was subsequently computed with the radiative transfer according to the primary transit geometry \citep{Ehrenreich2006}}. The absorption coefficients and wavelengths for the three helium metastable lines were taken from the NIST Atomic Spectra Database\footnote{\tt https://www.nist.gov/pml/atomic-spectra-database}. Doppler line shapes were assumed at the temperature of the helium model density. 
Additional broadening by the turbulent velocity were not included as we found that most of the absorption comes from radii smaller than the Roche lobe (4.22\,R$_{\rm P}$), where turbulence is not expected to be important. 
A mean velocity of the gas along the line of sight (towards the observer) was also included in order to account for {a possible bulk motion} of the absorbing gas. 

{Figure\,\ref{fig.transmission_spectrum} shows that} the observed absorption can be well reproduced for the helium triplet density obtained from that model for a temperature of 6000\,K and a mass-loss rate of {4.2$\times 10^{9}$\,g\,s$^{-1}$} (red curve). 
{The mean molecular weight obtained from the model for that fit is 0.76\,amu}. However, a degeneracy exists between the atmospheric temperature and $\dot{M}$ in the model, so that  $\dot{M}$ ranges between about 10$^{8}$ to 10$^{11}$\,g\,s$^{-1}$ for temperatures of 4500 to 11500\,K, respectively. 
{With all due caution, these evaporation rates -- assuming a 90\%\,H and 10\%\,He atmosphere -- are similar to previous estimations of hydrogen escape based on Ly$\alpha$ observations \citep[e.g.,][]{ Vidal-Madjar2003, Koskinen2010, Bourrier2013}, or energy-limited escape \citep{Sanz-Forcada2011, Czesla2017}, which is reasonable as the atmosphere is assumed to be hydrogen dominated.} 
In addition, the model shows that the weakest component of the triplet is expected at around 1$\sigma$ level, 
so its non-detection is consistent with the given data quality. More details on the modelling and on the temperature and mass-loss rates results will be given in a future paper.

\section{Discussion \label{sect.discussion}}
Strong absorption {of the upper atmosphere} has previously been detected in the transmission spectrum of HD\,209458\,b in Ly$\alpha$ \citep{Vidal-Madjar2003}, {atomic carbon and oxygen \citep{Vidal-Madjar2004}, and  magnesium \citep{Vidal-Madjar2013},} at the 5--10\% level. The He\,{\sc i} absorption presented in this work is significantly lower, as expected due to the low fraction of helium atoms in the excited state required to produce the absorption line {(Sect.\,\ref{subsect.activity}). However, the mass-loss rates derived from our He\,{\sc i} analysis agree with those obtained from Ly$\alpha$ (see Sect.\,\ref{subsect.modelling}).}

{Previous observations by \cite{Moutou2003} and \cite{Nortmann2018} provided upper limits on the detection of He\,{\sc i} for HD\,209458\,b, {which are consistent with the result presented here}}. Moutou et al. derived an upper limit of 0.5\% for a 3\,$\AA$ bandwidth. Assuming this bandwidth, our He\,{\sc i} absorption detection (0.91\,$\pm$\,0.10\%, FWHM$\sim$0.4\,$\AA$) corresponds to a value of 0.12\%.  
{In the case of Nortmann et al., the study of helium absorption was hampered by the poor data quality. The two datasets used in their {analyses} exhibit significantly lower signal-to-noise ratio than our observations. Their data were obtained with the same instrument, however, before an extensive intervention of the NIR channel in November 2016, which improved the thermal management of the channel \citep{Quirrenbach2018} and led to a significant improvement in the achievable data quality. 
In addition, the observational settings were not optimal, viz., the atmospheric dispersion corrector was not properly updated and calibration images were obtained during transit resulting in a $\sim$30\% loss of signal for one of the nights. Nonetheless, the data of both nights showed hints of absorption at the He\,{\sc i} position (see panels C and D on their Fig.\,S10), for which Nortmann et al. retrieved an upper limit of 0.84\% (90\% confidence level)}. 

If the helium signal were to be pursued with WFC3 on the {\it Hubble Space Telescope}, its spectral resolution of 98\,$\AA$ would result in a transmission signal of $\sim$10\,ppm. Even with NIRSPEC \citep{Dorner2016} on the soon to be launched {\it James Webb Space Telescope} ({\it JWST}), this signal will only be at the $\sim$300\,ppm level over one resolution element, at the highest resolving power {($\mathcal{R}$\,$\sim$\,2700)}. 

\subsection{Transmission line profile \label{subsect.line_profile}}

{One possible source of interference with the planetary He\,{\sc i} absorption could be the Rossiter-McLaughlin effect (RME). However, this effect can be neglected as the stellar He\,{\sc i} is very weak in HD\,209458. In fact, for similar targets where the stellar helium is stronger, the RME is estimated to be smaller than 0.1\% \citep[e.g.,][]{Nortmann2018, Salz2018}. In addition, the nearby Si\,{\sc i} line at $\sim$10827\,$\AA$ is almost nine times deeper than the He\,{\sc i} feature and leaves residuals in the average transmission spectrum of 0.2--0.4\%. Thus, we estimated that the maximum interference in the planetary He\,{\sc i} absorption caused by the RME should be smaller than 0.044\% (i.e., 20 times smaller than the measured transmission signal).}

Around mid-transit, we {noted} a possible additional absorption feature about 10--15\,km\,s$^{-1}$ blueward from the main absorption (Fig.\,\ref{fig.transmission_spectrum}). We measured in the HD\,209458 spectra several activity indicators related {to Ca\,{\sc ii} (infrared triplet), Na, and H$\alpha$} using the program {\sc Starmod} \citep{Montes2000}. There is no indication that HD\,209458 happened to be, {at any moment during the observations}, in a higher state of activity than reported in the literature \citep[e.g.,][]{Czesla2017}. {Thus, a possible activity impact on the transmission spectrum is unlikely to explain the additional absorption.}

{We cannot exclude that this particular absorption component is an artefact due to {an unsatisfactory correction} of an OH-doublet of sky emission lines at $\sim$10830\,$\AA$ (Fig.\,\ref{fig.master_spectrum}), but no such effect is seen for the significantly brighter OH-emission line at 10832\,\AA.} Alternatively, the presence of bad pixels in the area, in particular on the blue side of the main He\,{\sc i} {component, could also be responsible of the extra absorption}. 
In addition to using the standard bad pixel mask, we developed a method to identify and correct bad pixels {more carefully} (see Sect.\,\ref{section.analysis}). {It did not significantly reduce the possible absorption feature on the blue side of the main helium signal either.} 

{If real, the signal can be fitted by an {extra} absorption component {in the average transmission spectrum} from which a bulk velocity shift of $-$13\,km\,s$^{-1}$ can be estimated {(Fig.\,\ref{fig.transmission_spectrum})}. This agrees with the expected velocity of the escaping atmosphere at the Roche lobe height \citep{Salz2016}. Although this suggests that it originates at very large altitudes, which could be in the outer layer of the thermosphere or even in the exosphere, a simple radiative transfer model is still valid to estimate the velocity and absorber amount of the feature and thus, tentatively fit the profile of the absorption signal. {However, this fit does not contain information on the temporal variation of the feature that could support its planetary origin. Further observations are needed to confirm this additional absorption.}}

\subsection{Stellar irradiation and He\,{\sc i} signal \label{subsect.activity}}
\cite{Nortmann2018} presented a  relation between the strength of the observed He\,{\sc i} absorption and the stellar {XUV (5--504\,\AA)} irradiation, similar to that previously observed in stellar coronae by \cite{Sanz-Forcada&Dupree2008}. The absorption line originates from neutral helium atoms in an excited metastable 2$^3$\,S state. The population of this level takes place after ionisation of {He\,{\sc i}} atoms by incoming irradiation from the host
star, followed by recombination in a cold environment. {This radiation is generated in the corona and transition region
of late type stars (late F, G, K and M), and it is directly related to
the level of activity, which in turn depends mainly on stellar rotation. Thus close-in gaseous planets around late type stars are prime targets to search for the He triplet, considering also that active stars will
likely produce higher levels of XUV irradiation.}

\begin{figure}
\centering
\includegraphics[angle= 0, width=0.49\textwidth,,trim={20 210 20 230}, clip, keepaspectratio]{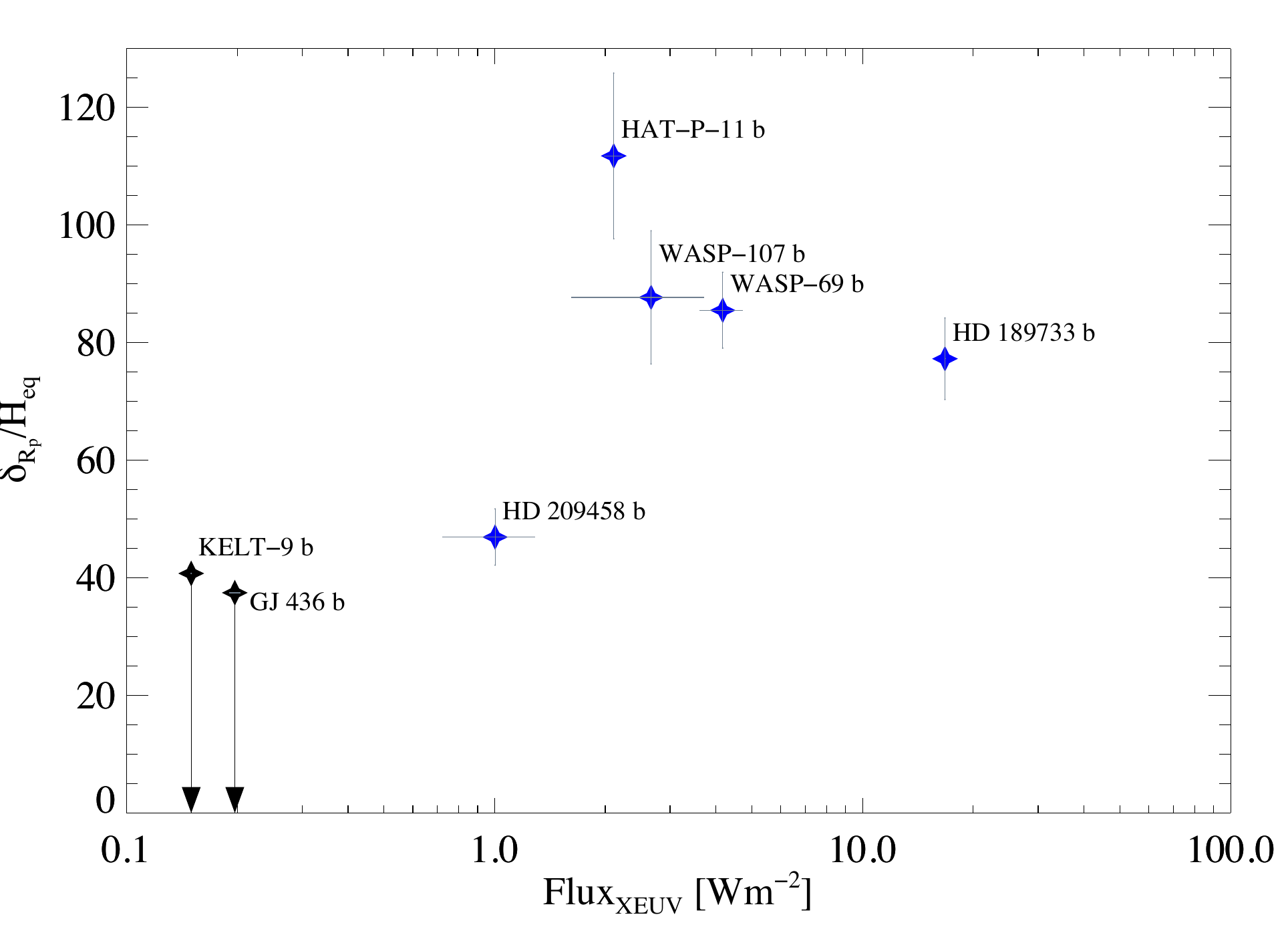}
\caption{\label{fig.irradiation} {He\,{\sc i} transmission signals currently detected (blue stars) and upper limits (black stars), as a function of the stellar irradiation below {504\,$\AA$} at the planet distance. We show the equivalent height of the He\,{\sc i} atmosphere, $\delta_{Rp}$, normalised by {the} atmospheric scale height of the respective planet’s lower atmosphere, $H_{\rm eq}$}.}
\end{figure}

We place our new measurement of the He\,{\sc i} absorption in the transmission spectrum of HD\,209458\,b in context of this hypothesis. 
In Fig.\,\ref{fig.irradiation}, we updated Fig.\,4 presented in \citet{Nortmann2018}, which showed the empirical relation between the stellar irradiation and the {detectability} of the He\,{\sc i} signal. 
{The {Y axis} indicates the equivalent height of the helium signal ($\delta_{Rp}$) normalised by the atmospheric scale height ($H_{eq}$). 
The values were computed using the data provided {by} \cite{Nortmann2018} and references therein, except for HD\,209458\,b (Table\,\ref{table.parameters}, $\delta_{Rp}/H_{\rm eq}$\,=\,46.9\,$\pm$\,4.8), HAT-P-11\,b 
\citep[$\delta_{Rp}/H_{eq}$\,=\,103.4\,$\pm$\,4.8]{Allart2018, Bakos2010, Deming2011}, and WASP-107\,b \citep[$\delta_{Rp}/H_{eq}$\,=\,87.7\,$\pm$\,11.3]{Allart2019, Anderson2017}. 
The X-ray and EUV (0.5--50.4\,nm) flux used for the {X axis} are from \citet{Nortmann2018}, except for HD\,209458\,b. 
Its value (see Table\,\ref{table.parameters}) was calculated using a modified version of the coronal model of \citet{Sanz-Forcada2011} and applied as explained {by} \citet{Nortmann2018}. The coronal model was updated after an improved fit of the summed {\it XMM}/EPIC spectrum (S/N\,=\,3.2,
${\log T}$\,(K)\,=\,6.0--6.3, $\log EM$\,(cm$^{-3}$)\,=\,49.52$^{+0.22}_{-0.48}$, {see also \citealt{Czesla2017}}) using the same spectra {as} in \cite{Sanz-Forcada2011}. The coronal model was extended to cooler temperatures using UV line fluxes from \citet{France2010}. 
Because of the lack of X-ray information for KELT-9\,b, the corresponding value is indicated as a lower limit.}
HD\,209458\,b has so far the weakest signal detected, and {it} is also the planet that receives the least XUV flux from its host star {--a relatively low-active G star}. Therefore, this measurement is in line with the {suggested trend and relevant to anchor the suspected activity relation at lower irradiation levels.}


{Although the distribution of all the signals in Fig.\,\ref{fig.irradiation} is consistent with a dependence on the XUV irradiation level, this may not be the only factor. \cite{Oklopcic2019} modelled the strength of He\,{\sc i} absorption in irradiated planetary atmospheres depending on the spectral type of the hosting star. They suggested that the ratio between extreme- and mid-UV irradiation fluxes determine the amplitude of this absorption, which particularly favours K stars. It remains to be investigated to what extent the mid-UV flux of HD\,209458 is responsible for the weaker helium absorption compared to the other four detections in planets orbiting K stars.}


\section{Conclusions \label{sect.conclusions}}
We present a {solid} detection of He\,{\sc i} $\lambda$\,10830\,$\AA$ in the transmission spectrum of the hot Jupiter HD\,209458\,b at a level of $0.91\,\pm\,0.10\,\%$. It concludes a search for He\,{\sc i} in this planet atmosphere for over a decade \citep{Seager&Sasselov2000,Moutou2003,Nortmann2018}. The strength of the detection {is consistent with} the empirical relationship proposed between the helium signal and the host star activity \citep{Nortmann2018}. 

We tentatively detect additional absorption on the blue side of the main helium signal at {about} $-$13\,km\,s$^{-1}$. Although this could be due to cometary-tail like escape, we are not confident yet about the reliability of the feature. In addition, there is no evidence for pre- or post-transit absorption in the He\,{\sc i} triplet. {Our spectral detection is consistent with models of atmospheric escape with {total hydrogen and helium} mass-loss rates of 10$^{8}$--10$^{11}$\,g\,s$^{-1}$ depending on the assumed temperature of the upper atmosphere}.


\begin{acknowledgements}
We thank P. Molliere and A. Wyttenbach for the nice scientific discussions during the preparation of this publication. F.J.A.-F. and I.S. acknowledge funding from the European Research Council (ERC) under the European Union Horizon 2020 research and innovation programme under grant agreement No 694513.  CARMENES is funded by the German Max-Planck-Gesellschaft (MPG), the Spanish Consejo Superior de Investigaciones Cient{\'i}ficas (CSIC), the European Union through FEDER/ERF FICTS-2011-02 funds, and the members of the CARMENES Consortium (MaxPlanck-Institut f{\"u}r Astronomie, Instituto de Astrof{\'i}sica de Andaluc{\'i}a, Landessternwarte K{\"o}nigstuhl, Institut de Ciències de l'Espai, Insitut f{\"u}r Astrophysik G{\"o}ttingen, Universidad Complutense de Madrid, Th{\"u}ringer Landessternwarte Tautenburg, Instituto de Astrof{\'i}sica de Canarias, Hamburger Sternwarte, Centro de Astrobiolog{\'i}a and Centro Astron{\'o}mico Hispano-Alem{\'a}n), with additional contributions by the Spanish Ministry of Economy, the German Science Foundation through the Major Research Instrumentation Programme and DFG Research Unit FOR2544 “Blue Planets around Red Stars”, the Klaus Tschira Stiftung, the states of Baden-W{\"u}rttemberg and Niedersachsen, and by the Junta de Andaluc{\'i}a. Financial support was also provided by the {Universidad Complutense de Madrid}, the Comunidad Aut\'onoma de Madrid, the Spanish Ministerios de Ciencia e Innovaci\'on and of Econom\'ia y Competitividad, the State Agency for Research of the Spanish MCIU through the ``Center of Excellence Severo Ochoa" and Science \& Technology Facility Council Consolidated, and the Fondo Social Europeo. The corresponding funding grants are: 
ESP2014--54362--P, 
ESP2014--54062--R, 
AYA2015-69350--C3--2--P,
BES--2015--074542,
AYA2016-79425--C3--1/2/3--P,
ESP2016--76076--R, 
ESP2017--87143--R,
SEV--2017--0709,
ST/P000592/1. 
Based on observations collected at the Centro Astron\'omico Hispano Alem\'an (CAHA) at Calar Alto, operated jointly by the Max--Planck Institut f\"ur Astronomie and the Instituto de Astrof\'{\i}sica de Andaluc\'{\i}a. We thank the anonymous referee for their insightful comments, which contributed to improve the quality of the manuscript.
\end{acknowledgements}


\begin{thebibliography}{}
\bibitem[Allart et al.(2018)]{Allart2018} Allart, R., Bourrier, V., Lovis, C., et al.\ 2018, Science, 362, 1384 
\bibitem[Allart et al.(2019)]{Allart2019} {Allart, R., Bourrier, V., Lovis, C., et al.\ 2019, \aap, 623, A58}
\bibitem[Anderson et al.(2017)]{Anderson2017} Anderson, D.~R., Collier Cameron, A., Delrez, L., et al.\ 2017, \aap, 604, A110 
\bibitem[Bakos et al.(2010)]{Bakos2010} Bakos, G.~{\'A}., Torres, G., P{\'a}l, A., et al.\ 2010, \apj, 710, 1724 
\bibitem[Ben-Jaffel(2007)]{Ben-Jaffel2007} Ben-Jaffel, L.\ 2007, \apjl, 671, L61 
\bibitem[Bourrier \& Lecavelier des Etangs(2013)]{Bourrier2013} Bourrier, V., \& Lecavelier des Etangs, A.\ 2013, \aap, 557, A124 
\bibitem[Bourrier et al.(2017)]{Bourrier2017} Bourrier, V., Ehrenreich, D., Wheatley, P.~J., et al.\ 2017, \aap, 599, L3 
\bibitem[Bourrier et al.(2018)]{Bourrier2018} {Bourrier, V., Lecavelier des Etangs, A., Ehrenreich, D., et al.\ 2018, \aap, 620, A147}
\bibitem[Brogi et al.(2017)]{Brogi2017} Brogi, M., Line, M., Bean, J., D{\'e}sert, J.-M., \& Schwarz, H.\ 2017, \apjl, 839, L2 
\bibitem[Caballero et al.(2016)]{Caballero2016} Caballero, J.~A., Gu{\`a}rdia, J., L{\'o}pez del Fresno, M., et al.\ 2016, SPIE, 9910, 0E 
\bibitem[Charbonneau et al.(2000)]{Charbonneau2000} Charbonneau, D., Brown, T.~M., Latham, D.~W., \& Mayor, M.\ 2000, \apjl, 529, L45 
\bibitem[Charbonneau et al.(2002)]{Charbonneau2002} Charbonneau, D., Brown, T.~M., Noyes, R.~W., \& Gilliland, R.~L.\ 2002, \apj, 568, 377
\bibitem[Crossfield et al.(2012)]{Crossfield2012} Crossfield, I.~J.~M., Knutson, H., Fortney, J., et al.\ 2012, \apj, 752, 81 
\bibitem[Czesla et al.(2017)]{Czesla2017} Czesla, S., Salz, M., Schneider, P.~C., Mittag, M., \& Schmitt, J.~H.~M.~M.\ 2017, \aap, 607, A101 
\bibitem[Deming et al.(2005a)]{Deming2005a} Deming, D., Seager, S., Richardson, L.~J., \& Harrington, J.\ 2005, \nat, 434, 740
\bibitem[Deming et al.(2005b)]{Deming2005b} Deming, D., Brown, T.~M., Charbonneau, D., Harrington, J., \& Richardson, L.~J.\ 2005, \apj, 622, 1149 
\bibitem[Deming et al.(2011)]{Deming2011} Deming, D., Sada, P.~V., Jackson, B., et al.\ 2011, \apj, 740, 33 
\bibitem[Dorner et al.(2016)]{Dorner2016} Dorner, B., Giardino, G., Ferruit, P., et al.\ 2016, \aap, 592, A113 
\bibitem[Ehrenreich et al.(2006)]{Ehrenreich2006} Ehrenreich, D., Tinetti, G., Lecavelier des Etangs, A., Vidal-Madjar, A., \& Selsis, F.\ 2006, \aap, 448, 379
\bibitem[Ehrenreich et al.(2008)]{Ehrenreich2008} Ehrenreich, D., Lecavelier des Etangs, A., H{\'e}brard, G., et al.\ 2008, \aap, 483, 933
\bibitem[Ehrenreich et al.(2015)]{Ehrenreich2015} Ehrenreich, D., Bourrier, V., Wheatley, P.~J., et al.\ 2015, \nat, 522, 459 
\bibitem[France et al.(2010)]{France2010} France, K., Stocke, J.~T., Yang, H., et al.\ 2010, \apj, 712, 1277 
\bibitem[Gaia Collaboration et al.(2018)]{Gaia2018} Gaia Collaboration, Brown, A.~G.~A., Vallenari, A., et al.\ 2018, \aap, 616, A1
\bibitem[Hawker et al.(2018)]{Hawker2018} Hawker, G.~A., Madhusudhan, N., Cabot, S.~H.~C., \& Gandhi, S.\ 2018, \apjl, 863, L11 
\bibitem[Henry et al.(2000)]{Henry2000} Henry, G.~W., Marcy, G.~W., Butler, R.~P., \& Vogt, S.~S.\ 2000, \apjl, 529, L41 
\bibitem[H{\o}g et al.(2000)]{Hog2000} H{\o}g, E., Fabricius, C., Makarov, V.~V., et al.\ 2000, \aap, 355, L27.
\bibitem[Hoeijmakers et al.(2015)]{Hoeijmakers2015} Hoeijmakers, H.~J., de Kok, R.~J., Snellen, I.~A.~G., et al.\ 2015, \aap, 575, A20 
\bibitem[Kausch et al.(2015)]{Kausch2015} Kausch, W., Noll, S., Smette, A., et al.\ 2015, \aap, 576, A78 
\bibitem[Knutson et al.(2007)]{Knutson2007} Knutson, H.~A., Charbonneau, D., Noyes, R.~W., Brown, T.~M., \& Gilliland, R.~L.\ 2007, \apj, 655, 564
\bibitem[Knutson et al.(2008)]{Knutson2008} Knutson, H.~A., Charbonneau, D., Allen, L.~E., Burrows, A., \& Megeath, S.~T.\ 2008, \apj, 673, 526
\bibitem[Koskinen et al.(2010)]{Koskinen2010} Koskinen, T.~T., Yelle, R.~V., Lavvas, P., \& Lewis, N.~K.\ 2010, \apj, 723, 116 
\bibitem[Kulow et al.(2014)]{Kulow2014} Kulow, J.~R., France, K., Linsky, J., \& Loyd, R.~O.~P.\ 2014, \apj, 786, 132 
\bibitem[Lammer et al.(2008)]{Lammer2008} Lammer, H., Kasting, J.~F., Chassefi{\`e}re, E., et al.\ 2008, \ssr, 139, 399 
\bibitem[Lecavelier des Etangs et al.(2010)]{Lecavelier2010} Lecavelier des Etangs, A., Ehrenreich, D., Vidal-Madjar, A., et al.\ 2010, \aap, 514, A72
\bibitem[Lavie et al.(2017)]{Lavie2017} {Lavie, B., Ehrenreich, D., Bourrier, V., et al.\ 2017, \aap, 605, L7}
\bibitem[Mazeh et al.(2000)]{Mazeh2000} Mazeh, T., Naef, D., Torres, G., et al.\ 2000, \apjl, 532, L55
\bibitem[Mansfield et al.(2018)]{Mansfield2018} Mansfield, M., Bean, J.~L., Oklop{\v c}i{\'c}, A., et al.\ 2018, \apjl, 868, L34 
\bibitem[Montes et al.(2000)]{Montes2000} Montes, D., Fern{\'a}ndez-Figueroa, M.~J., De Castro, E., et al.\ 2000, \aaps, 146, 103
\bibitem[Moutou et al.(2003)]{Moutou2003} Moutou, C., Coustenis, A., Schneider, J., Queloz, D., \& Mayor, M.\ 2003, \aap, 405, 341 
\bibitem[Nortmann et al.(2018)]{Nortmann2018} Nortmann, L., Pall{\'e}, E., Salz, M., et al.\ 2018, Science, 362, 1388 
\bibitem[Oklop{\v c}i{\'c} \& Hirata(2018)]{Oklopcic&Hirata2018} Oklop{\v c}i{\'c}, A., \& Hirata, C.~M.\ 2018, \apjl, 855, L11 
\bibitem[Oklop{\v c}i{\'c}(2019)]{Oklopcic2019} Oklop{\v c}i{\'c}, A.\ 2019, \apj , submitted, arXiv:1903.02576 
\bibitem[Parker(1958)]{Parker1958} Parker, E.~N.\ 1958, \apj, 128, 664 
\bibitem[Quirrenbach et al.(2016)]{Quirrenbach2016} Quirrenbach, A., Amado, P.~J., Caballero, J.~A., et al.\ 2016, SPIE, 9908, 12
\bibitem[Quirrenbach et al.(2018)]{Quirrenbach2018} Quirrenbach, A., Amado, P.~J., Ribas, I., et al.\ 2018, SPIE, 10702, 0W 
\bibitem[Salz et al.(2016)]{Salz2016} Salz, M., Czesla, S., Schneider, P.~C., \& Schmitt, J.~H.~M.~M.\ 2016, \aap, 586, A75 
\bibitem[Salz et al.(2018)]{Salz2018} Salz, M., Czesla, S., Schneider, P.~C., et al.\ 2018, \aap, 620, A97 
\bibitem[Sanz-Forcada \& Dupree(2008)]{Sanz-Forcada&Dupree2008} Sanz-Forcada, J., \& Dupree, A.~K.\ 2008, \aap, 488, 715 
\bibitem[Sanz-Forcada et al.(2011)]{Sanz-Forcada2011} Sanz-Forcada, J., Micela, G., Ribas, I., et al.\ 2011, \aap, 532, A6
\bibitem[Schwarz et al.(2015)]{Schwarz2015} Schwarz, H., Brogi, M., de Kok, R., Birkby, J., \& Snellen, I.\ 2015, \aap, 576, A111 
\bibitem[Seager \& Sasselov(2000)]{Seager&Sasselov2000} Seager, S., \& Sasselov, D.~D.\ 2000, \apj, 537, 916 
\bibitem[Seifert et al.(2012)]{Seifert2012} Seifert, W., S{\'a}nchez Carrasco, M.~A., Xu, W., et al.\ 2012, SPIE, 8446, 33
\bibitem[Shulyak et al.(2019)]{Shulyak2019} Shulyak, D., Reiners, A., Nagel, E., et al.\ 2019, \aap, {626, A86} 
\bibitem[Smette et al.(2015)]{Smette2015} Smette, A., Sana, H., Noll, S., et al.\ 2015, \aap, 576, A77 
\bibitem[Snellen et al.(2008)]{Snellen2008} Snellen, I.~A.~G., Albrecht, S., de Mooij, E.~J.~W., \& Le Poole, R.~S.\ 2008, \aap, 487, 357
\bibitem[Snellen et al.(2010)]{Snellen2010} Snellen, I.~A.~G., de Kok, R.~J., de Mooij, E.~J.~W., \& Albrecht, S.\ 2010, \nat, 465, 1049
\bibitem[Spake et al.(2018)]{Spake2018} Spake, J.~J., Sing, D.~K., Evans, T.~M., et al.\ 2018, \nat, 557, 68 
\bibitem[Skrutskie et al.(2006)]{Skrutskie2006} Skrutskie, M.~F., Cutri, R.~M., Stiening, R., et al.\ 2006, \aj, 131, 1163 
\bibitem[Swain et al.(2009)]{Swain2009} Swain, M.~R., Tinetti, G., Vasisht, G., et al.\ 2009, \apj, 704, 1616 
\bibitem[Torres et al.(2008)]{Torres2008} Torres, G., Winn, J.~N., \& Holman, M.~J.\ 2008, \apj, 677, 1324 
\bibitem[Vidal-Madjar et al.(2003)]{Vidal-Madjar2003} Vidal-Madjar, A., Lecavelier des Etangs, A., D{\'e}sert, J.-M., et al.\ 2003, \nat, 422, 143 
\bibitem[Vidal-Madjar et al.(2004)]{Vidal-Madjar2004} Vidal-Madjar, A., D{\'e}sert, J.-M., Lecavelier des Etangs, A., et al.\ 2004, \apjl, 604, L69 
\bibitem[Vidal-Madjar et al.(2013)]{Vidal-Madjar2013} {Vidal-Madjar, A., Huitson, C.~M., Bourrier, V., et al.\ 2013, \aap, 560, A54}
\bibitem[Wright et al.(2011)]{Wright2011} Wright, J.~T., Fakhouri, O., Marcy, G.~W., et al.\ 2011, \pasp, 123, 412 
\bibitem[Zechmeister et al.(2014)]{Zechmeister2014} Zechmeister, M., Anglada-Escud{\'e}, G., \& Reiners, A.\ 2014, \aap, 561, A59 

\end{thebibliography}
\end{document}